\documentclass[useAMS,usenatbib,longabstract]{mn2e}
\usepackage{graphicx,fleqn}
\usepackage{deluxetable}
\usepackage{longtable}

\def \apj{ApJ}
\def \apjl{ApJL}

\def\ltsim{\raisebox{-.5ex}{$\;\stackrel{<}{\sim}\;$}}
\def\gtsim{\raisebox{-.5ex}{$\;\stackrel{>}{\sim}\;$}}

\title[GRB 130427A]{The 80 Ms follow-up of the X-ray afterglow of GRB 130427A challenges the standard forward shock model.}
\author[]{M. De Pasquale$^{1,2,3}$, M. J. Page$^{1}$, D. A. Kann$^{4}$, S. R. Oates$^{1,5}$, S. Schulze$^{6,7}$,
\newauthor B. Zhang$^{8}$, Z. Cano$^{9}$, B. Gendre$^{10,11,16}$, D. Malesani$^{12}$, A. Rossi$^{13}$ E. Troja$^{14}$, 
\newauthor L. Piro$^{15}$, M. Bo{\"e}r$^{16}$, G. Stratta$^{17}$, N. Gehrels$^{14}$.\\
$^{1}$Mullard Space Science Laboratory - University College London, Holmbury Rd, Dorking, RH5 6NT United Kingdom.\\
E-mail: m.depasquale@ucl.ac.uk\\
$^{2}$Istituto Astrofisica Spaziale e Fisica Cosmica Palermo (INAF), Palermo, Via U. La Malfa 153, 90146 Palermo, Italy.\\
$^3$Istituto Euro Mediterraneo di Scienza e Tecnologia, Palermo, Italy.\\
$^4$Th{\"u}ringer Landessternwarte Tautenburg, Sternwarte 5, 07778 Tautenburg, Germany.\\
$^5$Instituto de Astrofisica de Andalucia (CSIC), Glorieta de Astronomia, E-18008 Granada, Spain.\\
$^6$Millennium Institute of Astrophysics, Vicu\~{n}a Mackenna 4860, 7820436 Macul, Santiago, Chile.\\ 
$^7$Instituto de Astrof\'isica, Facultad de F\'isica, Pontificia Universidad Cat\'olica de Chile, Vicu\~{n}a Mackenna 4860\\
7820436 Macul, Santiago, Chile.\\ 
$^8$Department of Physics and Astronomy, University of Nevada Las Vegas, Las Vegas, USA.\\
$^9$Centre for Astrophysics and Cosmology, Science Institute, University of Iceland, 107 Reykjavik, Iceland.\\
$^{10}$University of Virgin Islands, 2 John Brewer's Bay, 00802, St Thomas, USA V.I..\\
$^{11}$Etelman Observatory, St Thomas, USA V.I..\\
$^{12}$Dark Cosmology Centre, Niels Bohr Institutet, University of Copenhagen, Juliane Maries Vej 30, 2100 Copenhagen \O, Denmark.\\
$^{13}$Istituto Astrofisica Spaziale e Fisica Cosmica di Bologna (INAF), Via P. Gobetti 101, 40129 Bologna, Italy.\\ 
$^{14}$NASA Goddard Space Flight Center, Greenbelt, MD 20771, USA.\\
$^{15}$Institute of Space Astrophysics and Planetology, Via F. Del Cavaliere 100, Rome, 00133, Italy.\\
$^{16}$CNRS-ARTEMIS, Boulevard de l'Observatoire, CS 34229, 06304 Nice Cedex 4, France.\\
$^{17}$Department of Physics, University of Urbino, V. S. Chiara 27, Urbino, 61029, Italy.}

\begin{document}

\date{Accepted ... Received ...}

\maketitle

\label{firstpage}

\begin{abstract}
GRB 130427A was the brightest gamma-ray burst detected in the last 30 years. With an equivalent isotropic energy output of $8.5\times10^{53}$~erg and redshift $z=0.34$, it uniquely combined very high energetics with a relative proximity to Earth. As a consequence, its X-ray afterglow has been detected by sensitive X-ray observatories such as {\it XMM-Newton} and {\it Chandra} for a record-breaking baseline longer than 80 million seconds. We present the X-ray light-curve of this event over such an interval. The light-curve shows a simple power-law decay with a slope $\alpha = 1.309 \pm 0.007$ over more than three decades in time (47 ks - 83 Ms). We discuss the consequences of this result for a few models proposed so far to interpret GRB 130427A, and more in general the significance of this outcome in the context of the standard forward shock model. We find that this model has difficulty in explaining our data, in both cases of constant density and stellar-wind circumburst media, and requires far-fetched values for the physical parameters involved.


\end{abstract}

\begin{keywords}
Gamma-Ray Burst, general -- Gamma-Ray Burst, individual (GRB 130427A) -- relativistic outflows -- stellar wind
\end{keywords}

\section{Introduction} Gamma-ray Bursts (GRBs) are the most luminous explosions in the Universe (see Kumar \& Zhang 2015 for a recent review). In this article, we focus on GRB 130427A, the most intense long GRB \citep{kou93} ever detected by the {\it Swift} Burst Alert Telescope (BAT; \citealp{barthelmy04}) instrument. This event had an equivalent isotropic energy release of $8.5\times10^{53}$~erg at a redshift $z=0.34$ (\citealp{per14}; henceforth P14). The average isotropic energy release of {\it Swift} bursts is a few $\times~10^{52}$~erg. Events close to and above $10^{54}$~erg constitute only a few percent of all bursts \citep{kocevski2008,kann2010}; the average redshift of a GRB is $z=2.2$ (Jakobsson et al. 2012, \citealp{kru15}), while the percentile of GRBs at redshift $z<0.34$ is only $\sim 4\%$.   Thus, it is not surprising that events that release energy $\gtsim 10^{54}$ erg are usually found when taking into account very large volumes and large redshifts. The closest GRB with energetics comparable to GRB 130427A is 111209A, which produced $5.8\times10^{53}$~erg at redshift $0.67$ \citep{gendre2013,greiner2015}. The closest GRBs with energetic output higher than GRB 130427A are 080319B and 110918A \citep{cenko2010,fredericks2013}, which occurred at $z\simeq1$. \citet{cenko2010,cenko2011} examined the GRBs with the highest energetics detected by {\it Swift} and {\it Fermi} till 2010, which were objects at redshift of 1 or substantially higher.

At redshift equal to or lower than that of GRB 130427A, we typically detect low-luminosity GRBs (ll-GRBs; see \citealp{bromberg2011}) or transition events between ll-GRBs and long GRBs \citep{schulze2015}. ll-GRBs are a class of objects different from long GRBs. The main dissimilarity is their prompt energetics and luminosities, which are 2-4 orders of magnitudes lower than those of long GRBs. In addition, ll-GRBs often have lower peak energies and smoother prompt emission light-curves. Taken together, these differences imply that the physical phenomena in ll-GRBs are rather different from those that occur in long GRBs. The reader is referred to Fig. 1 of P14, which shows the isotropic energy releases vs redshift for pre-{\it Swift}, {\it Swift}, and {\it Fermi} long-duration GRB, to put GRB 130427 in the context of the energy vs distance distribution for known bursts.

GRB 130427A undoubtedly represents a very rare occurrence and,  given its proximity, it has enabled the GRB community to study the properties of the radiation mechanism and of the circumburst medium of very energetic GRBs in an unprecedented fashion. Unsurprisingly, a large body of literature on this GRB has already been written. Some articles deal with high-energy observations of the prompt emission (Ackermann et al. 2014, Preece et al. 2014), others focus on the X-ray (\citealp{kou13}, henceforth K13), optical \citep{vestrand14}, and radio afterglows (\citealp{laskar13,vdh14}, L13 and VA14), yet others present broadband afterglow modeling (P14 and \citealp{mas14}, M14), or study of the associated SN 2013cq \citep{xu13,melandri2014}.

However, this literature is based on observations taken up to $\simeq100$ days after the GRB trigger. The high energetics and low redshift of GRB 130427A have enabled X-ray observations to be obtained over a much longer and unprecedented time scale after the initial trigger, which allows us to test the currently accepted models put forward to describe this event.
In this article, we present X-ray observations of GRB 130427A performed up to 83~Ms after the trigger by {\it XMM-Newton} and {\it Chandra}. Even the latest observation resulted in a significant detection. To our knowledge, this is the longest time span over which the X-ray afterglow of a long GRB has been studied, and the 83~Ms datapoint represents the latest detection of a GRB X-ray afterglow. \cite{kou04} studied the field of GRB 980425A 3.5 years (i.e. 111~Ms) after the trigger, but this event belongs to the different class of ll-GRBs, and it is not clear whether the latest detection is due to the typical afterglow emission. Previously, the burst with the longest X-ray afterglow follow-up was GRB 060729 \citep{grupe10}; the afterglow of this burst was detected by {\em Chandra} 55.5~Ms after the trigger. Only in the radio band have follow up observations of GRBs occasionally extended further.

In Sect. \ref{oba}, we present the observations, the data reduction, and the results of our analyses. In Sect. 3 we model the X-ray afterglow of GRB 130427A, while in Sect. 4 we present our conclusions. We adopt the cosmological parameters determined by the {\it Planck} mission, i.e. $H_0 = 67.8$ km~s$^{-1}$~Mpc $\Omega_m = 0.31$, $\Omega_{\Lambda} =0.69$ \citep{planck}.
The afterglow emission is described by $F_{\nu} \propto t^{-\alpha} \nu^{-\beta}$, where $t$ is the time from trigger, $\nu$ the frequency, $\alpha$ and $\beta$ are the decay and spectral indices, respectively. Errors are reported at 68\% confidence level (C.L.) unless otherwise specified.





\section{Observations and Analysis of X-ray data}\label{oba}
\subsection{{\it Swift}-XRT observations}
The {\it Swift} X-ray Telescope (XRT; \citealp{burrows05}) began observing GRB~130427A (M14) less than 200~s after the start of the prompt emission (indicated as $T_0$), and continued to monitor
the source for $\simeq 15.8$~Ms ($\simeq180$~days). The XRT count-rate light-curve was obtained  from the UK {\em Swift} Science Data Centre on-line light-curve repository \citep{evans07, evans09}, and binned to a uniform $\Delta T / T = 0.05$. 

\subsection{{\it XMM-Newton} observations}\label{xmm}
GRB~130427A was observed with {\em XMM-Newton} (PI: De Pasquale) at seven times: 2013 May 13 ($T_0+1.4$~Ms), 2013 June 20, 2013 ($T_0+4.7$~Ms), 2013 November 14 and November 16 ($T_0+17.4$ and $T_0+17.6$~Ms respectively), 2015 May 31 ($T_0+66.1$~Ms) and December 12 and 24 ($T_0+82.9$~Ms and $T_0+84.0$ ~Ms respectively). The {\em XMM-Newton} data were reduced using the {\sc science analysis system (sas)} version 14.0. Periods of high background were identified in full-field light-curves of events with energy $>5$~keV, and excluded from the analysis. Table \ref{tab:xmmexposures} gives the exposure times for each observation after periods of high background have been excluded. At each epoch spectra were extracted from the MOS and pn data using a circular aperture centred on the source. The radius of the source aperture was chosen according to the brightness of the source: radii of 40 and 15 arcsec were chosen for the 2013 May 13 and 2013 June 20 data respectively.  For subsequent observations an aperture of radius 7 arcsec was chosen, the small aperture was necessitated by the presence of a nearby source which would otherwise contaminate the extracted spectrum of GRB~130427A. None of the spectra are affected by photon pile-up. The two observations taken in November 2013 are separated by only
$\Delta T /T \simeq 0.01$, so these data were combined to form a single point in the light-curve. Similarly, the two observations taken in December 2015 are separated by $\Delta T / T \simeq 0.01$, and were combined to form one point in the light-curve.

For each epoch the MOS and PN spectra were combined as described in \citet{page03}. 

\begin{table}
\caption{{\em XMM-Newton} EPIC exposure details. The exposure times are given after periods of high background have been removed.}
\label{tab:xmmexposures}
\begin{tabular}{ccccc}
OBSID&Date&\multicolumn{3}{c}{Exposure time (ks)}\\
           &         &MOS1&MOS2&pn\\
\hline
0693380301&2013 May 13&28.3&28.3&24.0\\
0693380501&2013 June 20&16.1&16.1&13.1\\
0727960701&2013 Nov 14&25.1&24.9&13.5\\
0727960801&2013 Nov 16&18.0&18.3&13.0\\
0764850201&2015 May 31&60.0&60.0&51.0\\
0764850301&2015 Dec 12&33.5&32.5&23.2\\
0764850401&2015 Dec 24&14.1&13.7&9.4\\
\hline
\end{tabular}
\end{table}

\subsection{{\it Chandra} observations}\label{Chandra}
{\em Chandra} observed GRB~130427A on 2014 February 11 and 2014 June
21 (PI: Fruchter, Obs ID 14885, 14886), which correspond to $T_0+25.1$~Ms and $T_0+36.3$~Ms, respectively. We used these publicly available data from the \textit{Chandra} archive to build our light-curve. The exposure times of the two observations were 19.8 ks and 34.6 ks. In both observations the target was placed on the ACIS S3 chip. The source photometry was measured using the task {\sc
  wavdetect} in the reduction package {\sc ciao} version 4.8.


\subsection{Building the X-ray light-curve}
In the subsections above we summarized the observations of the X-ray instruments and data reduction. The resulting flux of each {\em XMM-Newton} and {\em Chandra} observation is shown in Table 2.  We now describe how we combined these different datasets into a homogeneous flux light-curve.

We assumed the {\em XMM-Newton} derived spectral parameters (see Sect. \ref{results}) to
translate the measurements from {\em Swift} XRT, {\em Chandra} and {\em
  XMM-Newton} to 0.3-10~keV flux units in a consistent fashion. For
{\em XMM-Newton} the fluxes were derived directly from the spectra
using {\sc xspec}. For {\em Swift} XRT the conversion factor from
count rate to flux was obtained with the {\sc portable multi-mission
interactive simulator (pimms}\footnote{https://heasarc.gsfc.nasa.gov/cgi \\
bin/Tools/w3pimms/w3pimms.pl}, \citealp{mukai93}); in particular, the difference between our conversion
factor and that found at UK Swift Science Data Centre at the University of Leicester
is only $0.35\%$. We also obtained a good match between the {\em XMM-Newton}
and {\it Swift} measurements when observations were simultaneous.
For {\em Chandra} the conversion factors from count rates to flux were derived using {\sc xspec} from response files generated for the two observations using the {\sc ciao} script {\sc specextract}. A 10 per cent
uncertainty was added to the errors on the {\em Chandra} and {\em
  XMM-Newton} fluxes to account for systematic calibration differences
between these instruments and {\em Swift} XRT
\citep{tsujimoto11}.






\begin{table*}
\begin{center}
\caption{Flux of GRB 130427A afterglow in the 0.3-10 keV band, as measured by {\it XMM-Newton} and {\it Chandra} observations presented in this work. Note: we have added a 10\% systematics error these quantities before building the composite light-curve shown in Figure 1 and fitting it. Note: the third and the last line show the mid-time for the {\em XMM-Newton} observations taken in November 2013 and December 2015. \label{tab:flux}}
\begin{tabular}{llllllllllll}
\hline
$T - T_0$ & Facility & Flux \\
 Ms & & $\times10^{-15}$ cgs & \\
\hline
& & & \\
1.38  & {\em XMM-Newton} &  $849.94  \pm 0.85$ \\
4.66  & {\em XMM-Newton} &   $165.62 \pm 4.97$ \\
17.50 & {\em XMM-Newton} &  $26.57  \pm 2.15$ \\
25.06 & {\em Chandra}         &  $16.74  \pm 3.16$ \\
36.31 & {\em Chandra}         &  $12.12  \pm 2.11$ \\
66.06 & {\em XMM-Newton} &  $5.00    \pm  0.81$ \\
83.45  & {\em XMM-Newton} &  $3.54    \pm  0.84$ \\
\end{tabular}
\end{center}
\end{table*}

\begin{figure*}\label{lc}
\begin{center}

\includegraphics[angle=00,scale=0.36]{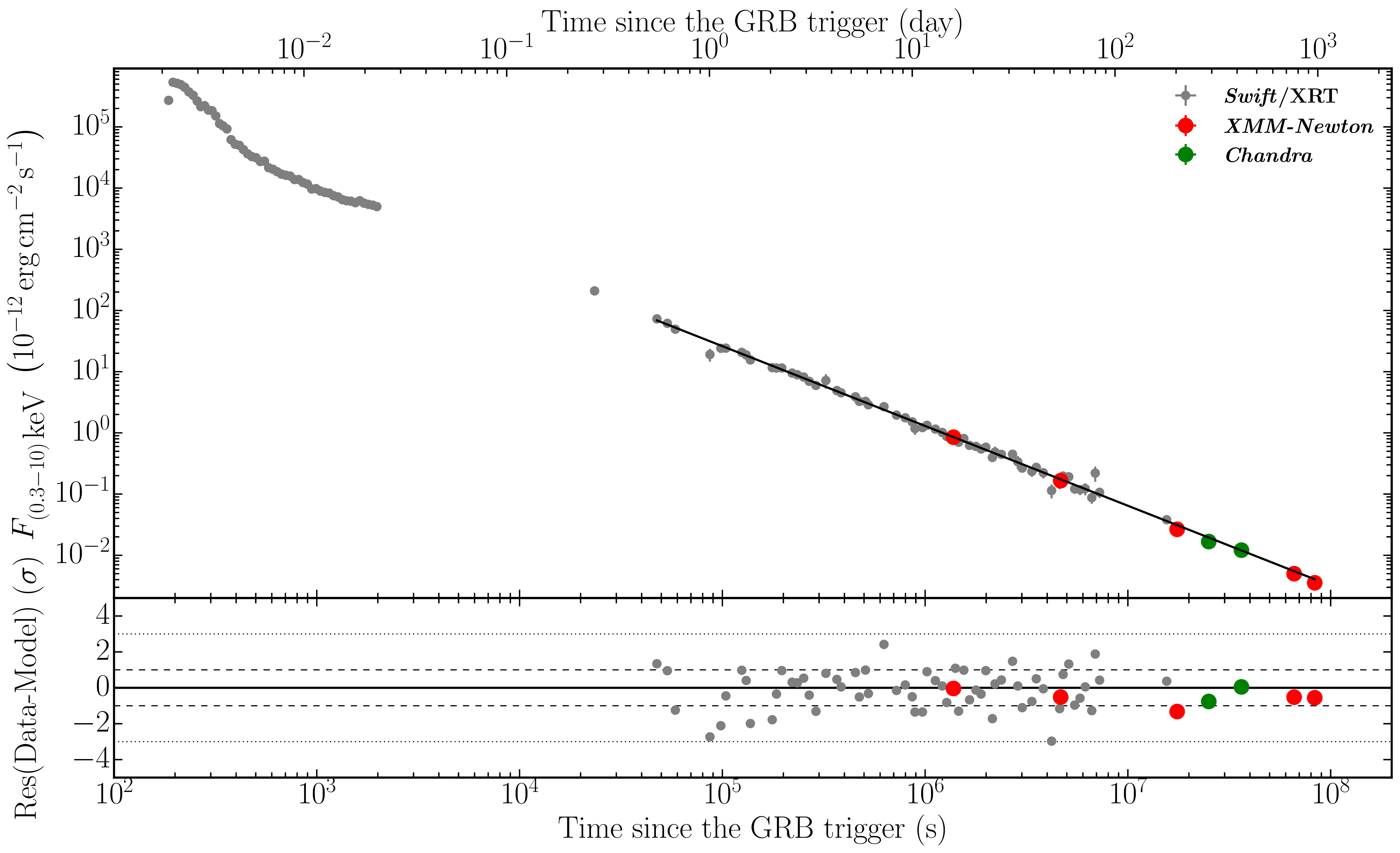}

\caption{X-ray light-curve of GRB 130427A. XRT, {\it Chandra} and {\it XMM-Newton} data are displayed in black, green and {\bf red} respectively. We superimpose the best fit, a simple power-law model with $\alpha = 1.309$ (see text for details).}

\end{center}
\end{figure*}

\subsection{Results}\label{results}
Among our observations, the {\it XMM-Newton} one on 2013 May 13 obtained the highest quality spectrum of GRB 130427A. We fitted the spectrum in {\sc xspec} version 11.0
\citep{arnaud96} with a power-law model, attenuated by a fixed
Galactic column density of $1.8\times 10^{20}$~cm$^{-2}$ and a second
photoelectric absorber at the redshift of the GRB. The best fit
power-law energy index is $\beta = 0.79\pm0.03$ and the best fit
host galaxy column density is ($5.5\pm0.6)\times
10^{20}$~cm$^{-2}$. These values are in excellent agreement with those
derived from the {\em Swift} XRT PC mode data, which are $\beta =0.72 \pm 0.04$ and N$_H = (6 \pm 1)\times 10^{20}$ cm$^{-2}$  (M14). The
other {\em XMM-Newton} observations yield spectra which are consistent 
with these values. 
We show the observed X-ray light-curve of GRB 130427A from the trigger to 83~Ms in Fig.~1. However, in our analysis we considered data from 47~ks (that is, after the 3rd {\it Swift} orbit). We decided to exclude prior data because we are interested in the late X-ray afterglow; {\em our discussion focuses on the consistency between models and late X-ray data.}

When fitting this X-ray light-curve with a simple power-law model, we obtain $\alpha = 1.309 \pm 0.007$. This fit model yields $\chi^2 = 75.8$ with 66 degrees of freedom (d.o.f.). The decay slope is similar to the previous measurements obtained over a smaller timescale. For example, M14, L13, P14 and K13 determined $\alpha = 1.35 \pm0.01$, $\alpha \simeq 1.35$, $\alpha = 1.35$, and $\alpha \simeq 1.281\pm0.004$ respectively, using data up to $\simeq 100$~days after the trigger. To test for the presence of any break after 47 ks, we fit our light-curve with a smoothly broken power-law model \citep{beu99}, with a smoothness parameter $n=2$. Empirically, we find that this smoothness parameter corresponds to a change of slope occurring over $\simeq2$ decades in time. The choice of such a smooth break is motivated by the findings that some jet breaks might occur over more than 1 decade in time \citep{granot2007}.
The fit with the Beurmann model gives $\chi^2 / {\rm dof}= 74.1 / 64$. According to the F-test, the probability $P$ of an improvement over the simple power-law model by chance is $P=0.48$.
We have also tested the presence of multiple breaks, by fitting the X-ray light-curve with a double broken power-law and a triple broken power-law models. The resulting fits yield $\chi^2 / {\rm dof} = 69.0/62$ and $\chi^2 / {\rm dof} = 64.1/60$ respectively. According to the F-test, the probability $P$ of an improvement by chance over the simple power-law model are $P=0.20$ for the double power-law model and $P=0.11$ for the triple broken power-law model. Every probability calculated by means of the F-test is high, which leads us to conclude that a break or multiple breaks are not required by the light-curve.



 
\section{Discussion}\label{discussion}
We will now explore the durability of several models proposed in the literature for GRB 130427A, which were built on the basis of data up to $\simeq100$~days after the trigger. We will check whether these models can still hold when we include new data gathered over much longer timescales. We will not examine models that do not assume the emission mechanisms typically invoked for afterglows (for example \citealp{dado2016}) or heavily modify them (e.g. \citealp{vurm2015}). The reader is referred to the aforementioned articles for a different approach.

The standard forward shock (FS; for a review, see \citealp{gao13}) model for GRBs predicts several possible phenomena that leave their signatures on the late X-ray afterglow. The very long temporal baseline of GRB 130427A observations provides us with an ideal opportunity to look for these signatures. Among the phenomena of interest to us, we have the so-called jet break \citep{racusin09} and the change of the density profile of the circumburst medium. These features may present themselves in different ways depending on the density profile of the medium; in this respect, different authors have used diverse density profiles. For example, P14, L13, VA14 and K13 modeled the afterglow assuming that the medium has a decreasing, stellar-wind density profile, while M14 employed a constant density, interstellar-like medium (ISM). We will discuss the proposed models according to the density profile they employ. However, in jet break models where ejecta are expanding laterally \citep{sari99}, the post-break decay slope is independent of the density profile. Thus, we first estimated a possible break time for such a jet\footnote{We are aware that recent simulations do not show sideways expansion (e.g. \citealp{granot2007}). However, the decay slopes predicted in this scenario still provide a reasonable fit for the post-jet break slopes.}.
The decay slope after such a jet break becomes $\alpha=p$ for any $\nu > \nu_{\rm m},\nu_{\rm a}$ (where $\nu_{\rm m}$ and $\nu_{\rm a}$ are the synchrotron peak and self-absorption frequency, respectively) and if there is no evolution in the values of the physical parameters in the models, such as the energy of the ejecta and the fractions of the shock energy given to the radiating electrons and magnetic field, and the index $p$ of the power-law energy distribution of the radiating electrons. Previous modeling of the afterglow emission of GRB 130427A derived values of $p$ quite close to each other. They range from $p=2.1$ (VA14) to $p=2.34$ (K13). While the fit of the X-ray light-curve does not require the addition of a break to a steeper decay (see previous Section), we can still derive a $95\%$ C.L. lower limit on the epoch of a jet break with post-break slope slightly steeper than 2, as we describe below. We refitted the light-curve with the smoothly broken power-law model, freezing the late decay slope value to $2.2$. We then used different values for the break time, moving it backwards until we found $\Delta\chi^2 = 2.7$ with respect to the best fit.

Following this method, we found $t_{\rm jet} \stackrel{>}{\sim} 61$~Ms. Such a lower limit on the jet break time translates into a lower limit on the beaming angle of the outflow $\theta_{\rm jet}$ and, consequently, on the beaming-corrected energetics. The exact value of $\theta_{\rm jet}$ depends however on the assumptions on the density of the circumburst medium and the kinetic energy of the relativistic outflow (see e.g. \citealp{cenko2011}). We will consider the effects of the lower limit of the break time on the beaming angle and on the energetics in the next section.



\subsection{Stellar wind density profile}\label{wind_section} 
\subsubsection{Energetics}
P14 and L13 used a free stellar-wind profile for the density $\rho (r) = A\,r^{-2}$, where $r$ is the distance from the centre of the explosion. In such a condition, the opening angle of the ejecta $\theta_{\rm jet}$ is \citep{chl00}:

\begin{equation}\label{a}
\theta_{\rm jet} = 0.11 \left(\frac{t_{{\rm jet,d}}}{1+z}\right)^{1/4} \left(\frac{E_{{\rm K,iso,53}}}{A_{\star}}\right)^{-1/4}\, \rm{rad}
\end{equation} 

where $E_{\rm K,iso}$ is the kinetic energy of the relativistic ejecta assuming isotropy, $A_{\star} =  A/(5\times10^{11} {\rm g~cm^{-1}})$ is the normalization constant for the wind density\footnote{$5\times10^{11}$ g cm$^{-1}$ corresponds to a mass lost rate of $10^{-5}$ $M_{\odot}$ yr$^{-1}$ with a wind speed $v_{\rm wind} = 10^{8}$ cm s$^{-1}$. These values are typical for Wolf-Rayet stars, which are thought to be the progenitors of long duration GRBs like 130427A \citep{woosley06}.}, and we use the convention $Q_x = Q / 10^x$ in cgs units. L13 found $E_{{\rm K,iso}} \simeq 7\times10^{52}$~erg and a very small $A_{\star}\simeq3\times10^{-3}$, while P14 obtained $E_{{\rm K,iso}} \simeq 3\times10^{53}$~erg and the same\footnote{P14 actually find a range of possible values for these parameters, but this range is not very wide, and we adopt the median.} $A_{\star}$ as L13. Using the values of the energetics and wind density presented in the papers above, for $t_{\rm jet}>61$~Ms we find lower limits for the beaming angles of $\gtsim0.09$~rad (P14), $\gtsim0.13$~rad (L13). If we correct the kinetic energy and the energy radiated during the $\gamma$-ray prompt emission phase for the corresponding beaming factors, we obtain total, beamed-corrected energetics (kinetic + prompt) of  $\gtsim5.1\times10^{51}$~erg (P14), $\gtsim8.4\times10^{51}$~erg (L13). 
If we assume that no jet break has occurred at all throughout 83~Ms, then the lower limits on the beaming angles and corresponding beamed-corrected energy become 0.10~rad and $5.9\times10^{51}$~erg (P14), and 0.15~rad and $9.8\times10^{51}$~erg (L13). The lower limit for the energetics in the case of P14 is large but not extreme. However, the lower limit in the case of L13 would place 130427A close to the ``hyper-energetic" bursts (see \citealp{cenko2011,ackermann13,martin14}; note, however, that \citealp{cenko2011} derive the opening angle using a coefficient of Eq. \ref{a} different from ours). 

   

\subsubsection{The stellar bubble in the standard free stellar wind}

The fact that we do not see signatures of a transition between stellar wind and interstellar medium, which would be seen as a change in the decay slope, bears other interesting consequences. In the modeling of P14 and L13, the X-ray frequency is at or just below the synchrotron cooling frequency $\nu_{\rm c}$. This assumption entails that, if the ejecta transition from a medium with decreasing density, such as stellar wind, into one with a constant density, such as the interstellar medium, the observer should see a flattening of the light-curve. The decay slopes predicted by the FS model are $\alpha =3/4\,p -3/4 $ and $\alpha =3/4 \, p -1/4$ for $\nu< \nu_{\rm c}$ in the constant density and wind medium, respectively. Thus, in the case of the modeling of L13 and P14, which used a value of $p$ very close to 2.2, one would expect a new decay slope of $\alpha =0.9$. With the same method used to determine a lower limit for a jet break, we find a 95\% C.L. lower limit of 48~Ms for any change of slope of the light-curve to a shallow $\alpha=0.9$. In a stellar wind environment, the radius reached by the ejecta at a certain time $t$ is \citep{chl00}:

\begin{equation}\label{radius}
R = 1.7 \times 10^{19} E_{\rm K,iso,54}^{1/2}  A_{\star,-1}^{-1/2} \left(\frac{t_6}{1+z}\right) ^{1/2} {\rm cm}
\end{equation}
For the values of $A$ determined by L13 and P14, the corresponding radii at $t\gtsim48$~Ms are $\gtsim1.6\times10^{20}$~cm (about 50 pc) and $3.3\times10^{20}$~cm (about 105~pc), respectively. Thus, the GRB progenitor must have carved a region, i.e. a ``stellar wind bubble", that extends for several tens of parsecs. It is still unclear why some GRBs seem to occur in large bubbles, while others are found to have a constant density medium very close to the centre of explosion \citep{schulze11}. However, in the case of the wind models of GRB 130427A proposed by L13 and P14, the size of the bubble is larger than any other determined in the literature ({\em ibid.}, \citealp{grupe10}). For this to happen, the density, $n_0$ cm$^{-3}$, of the pre-existing medium must have been very small. Following \cite{fryer2006} and reference therein, the radius of the termination shock, i.e. the region where the density profile is $\rho(r) \propto r^{-2}$, is $r_{t} \simeq  \dot{M_{-5}}^{1/3} n_{0,2}^{-1/2} $~pc, where $\dot{M_{-5}}$ is the mass loss in units of $10^{-5} M_{\odot}$ yr$^{-1}$, and it does not depend strongly on the wind velocity.

 Assuming a wind velocity $v_{wind} = 10^8$ cm s$^{-1}$, for the mass loss constrained by L13 and P14 and the radii determined above, we obtain $n_0 \ltsim 9\times10^{-4}$ (L13) and $n_0 \ltsim 2\times10^{-4}$ (P14). However, if we assume that no change of slope is present up to the end of our observations, 83~Ms, these densities will become even smaller: $n_0 \ltsim 5\times10^{-4}$ (L13) and $n_0 \ltsim 1.2\times10^{-4}$ (P14). In the case of the modeling put forward by P14, the density of the pre-existing medium is almost unrealistically small and unlikely to be found in the medium of any galaxy. We have not found examples of star forming regions in the Milky Way or Magellanic Clouds occurring in pre-existing environments of such low density. \citet{hunt2009} and \citet{peimbert2013} merged several observational data sets for Galactic and extragalactic H II regions, and found no densities below $\sim1$ particle cm$^{-3}$. We have explored the possibility that the GRB progenitor exploded in a very large bubble blown by a cluster of massive stars; some of these massive stars might have produced supernova events as well. In this case, the termination shock of a single star might not be the relevant parameter. Star clusters containing a large number of OB stars may produce ``superbubbles" within the surrounding medium with radii of $\sim100$~pc. According to recent numerical simulations (see e.g. \citealp{sharma2014,yadav2016}), the density profile inside these regions is roughly $\rho\propto r^{-2}$, as hypothesised in the case of GRB 130427A, if a star cluster contains more than $\sim10^4$~OB stars. However, the aforementioned simulations show that, if the number of OB stars is below the $\sim10^4$ threshold, the superbubbles are still formed but the density profile inside them does not decrease with radius. It is instead roughly constant with radius and then increases towards the edge\footnote{Basically, \citet{scalo01} found the same result analytically.}, as expected in individual supernova remnants \citep{vietri2008} and bubbles blown by single stars \citep{weaver77}.
 
It is not clear whether clusters with a sufficiently high number of massive OB stars to produce a $\rho\propto r^{-2}$ density profile have been found. According to \citet{beck2015}, the most massive young star clusters found in the local universe (including the Local Group) have mass up to $\sim10^6$ solar masses; they may not contain more than $\sim 10^3$ OB stars. A few star forming regions, found in the Local Group and beyond, are inside large voids that look like superbubbles. Their edges may form ``supershells", with radii that can reach as large as hundreds of parsecs \citep{warren2011}. However, there are very few measurements of the total density inside these regions and how it decreases with radius. The density of atomic hydrogen is actually found to {\it increase} with radius in the case of the very large voids studied by \citet{warren2011} in the Local Group.
 

 It might be possible that the progenitor of GRB 130427A exploded outside the host galaxy; this occurrence may explain the very low density of the pre-existing environment. However, this circumstance is unlikely, because the GRB site is spatially consistent with a star-forming region in the host galaxy \citep{levan2014a}, and while the column density determined from the X-ray absorption ($\simeq6\times10^{20}$~cm$^{-2}$) is rather low, it is not so low as to be consistent with an extra-galactic origin.

In the modeling by L13, the density of the medium in which GRB progenitor wind expanded may not be so extreme. However, we note (as other authors have done, e.g. see M14) that this model needs to convert a very large fraction of the initial energy of the ejecta into prompt $E_{\gamma}$ and initial afterglow emission.

According to L13, the kinetic energy of the ejecta of GRB 130427A was $\simeq 2\times10^{53}$~erg just after the end of the prompt emission. Afterward, the afterglow was in a period of so-called radiative cooling \citep{sari98}, which lasted up to $t - T_0 \simeq  1200$~s. During such a phase, the ejecta convert a significant fraction of their kinetic energy into radiation. According to L13, the ejecta lost $E_{\rm rad} \simeq 1.3\times10^{53}$~erg during the radiative cooling, i.e. $65\%$ of their energy. When the radiative cooling ended, and the adiabatic cooling (the typical condition of most afterglows) began, the ejecta kinetic energy was reduced to $7\times10^{52}$ erg. 
The problem with this scenario is that, when modeling the GRB afterglow, L13 find $\epsilon_{\rm e} \simeq 0.3$, which is too low to cause substantial radiative losses. L13 assume that the deceleration time, i.e. the time when the ejecta pile up a fraction $\Gamma^{-1}$ of their mass and produce the onset peak in the FS emission, is $T_{\rm dec} \simeq 200$~s. With the value of $\epsilon_{\rm e}$ above, over a time scale of $t / T_{\rm dec} \simeq 6$, the ejecta lose only few percent of their total energy (see Fig. 5 of \citealp{nava2013}). To achieve a loss of 65\% over $t / T_{\rm dec} \simeq 6$, one would need $\epsilon_{\rm e} \simeq 0.8-1$, regardless of ISM or wind environment. This value of $\epsilon_{\rm e}$ is extremely high and difficult to explain; furthermore, it is at odds with the modeling of GRB 130427A by L13.
One has therefore to assume that the afterglow ejecta had a kinetic energy of $E_{\rm K}  \simeq7\times10^{52}$~erg from the deceleration time. The efficiency of converting the initial energy into prompt emission is  $\eta = E_{\gamma} / (E_{\gamma} + E_{\rm K})$. For $E_{\gamma} = 8.5\times10^{53}$~erg, and $E_{\rm K} = 7\times10^{52}$~erg determined by L13, the emission mechanisms must have had an efficiency of $\eta = 92\%$, which poses a very serious difficulty for any dissipation and emission models. We also note that using a wider dataset, which includes early GeV emission (not studied by L13), other authors (e.g. P14) do not find evidence for radiative cooling. 
To achieve a more moderate and more reasonable efficiency $\eta$, one should have a larger kinetic energy of the outflow; but this would imply a larger radius for the wind bubble (see Eq.~\ref{radius}) and, as a consequence, a rather low density of the pre-existing environment.

In the argument at the beginning of this subsection we have assumed that, when the ejecta enter a constant density environment, the cooling frequency $\nu_{\rm c}$ will stay above the observed X-ray frequency. However, the value of $\nu_{\rm c}$ depends on the density of the environment and we do not know the density of the medium outside the free stellar wind region of GRB 130427A; thus we cannot exclude that the cooling frequency moves below the observ{\bf ed} X-ray band as the ejecta leave the free stellar wind region. In such a case, the X-ray decay slope would be larger by 0.25 than in the case of $\nu_{\rm x} < \nu_{\rm c}$, i.e. $\alpha = 1.15$. Our observations, however, still do not match this prediction. Thus, we still do not find evidence of a transition between stellar wind and constant density medium in our observations. 

To summarize, our late X-ray observations indicate that wind models with the standard $\rho \propto r^{-2}$ profile advocated for GRB 130427A require either an emission process which is extraordinarily efficient in converting the initial energy of the explosion into prompt $\gamma$-ray emission, or a very low density of the pre-existing media. This very low density medium is difficult to explain for the apparent location of GRB 130427A inside its host galaxy.

\subsubsection{Non-standard wind environments}
K13 proposed that GRB 130427A occurred in a non-canonical stellar wind environment, with a density profile $\rho(r) = A\,r^{-1.4}$. We first modeled the X-ray light-curve of GRB 130427A assuming this slope for the density profile of the medium, and using the formulae of van der Horst (2007; their Tab. 2.5) to derive the values of the peak synchrotron flux, $\nu_{\rm m}$ and $\nu_{\rm c}$. We assumed that the spectral index above the cooling frequency $\nu_{\rm c}$ is\footnote{See Table 1 of K13.} $\beta=1.17$ and thus $p=2.34$. We took $\nu_{\rm c} > 2.4\times10^{18} \rm{Hz} = {\rm~10~keV}$. We imposed that the value of the Compton Parameter is $Y \stackrel{<}{\sim} 0.2$ at 20~ks; this constraint avoids inverse Compton emission, which is not observed in the model of K13 at this epoch (indeed, K13 conclude that the whole emission from optical to GeV is consistent with being synchrotron). In this assumption, we took into account the Klein-Nishina correction (see \citealp{zhang07}). Finally, we imposed a flux density $F_{\nu}(\rm{1~keV}) \simeq 20~\mu$Jy at 20 ks, as K13 show.

We found that an isotropic kinetic energy of the ejecta  $E_{\rm{K,iso}} \simeq 9.9\times10^{53}$~erg and the value of the parameter $A \simeq 0.001$ g cm$^{-1.6}$, which corresponds to\footnote{note that this value of the normalization implies a very thin medium, as in the cases treated by other authors. At a distance $r=6\times10^{19}$~cm (i.e. $\simeq20$ pc) from the centre of the explosion, one would have $\rho \simeq 1.2\times10^{-7}$ cm$^{-3}$, which compares with $\rho \simeq 2.5\times10^{-7}$ cm$^{-3}$ predicted for the models of L13 and P14.} $6\times10^{20}$ cm$^{-1.6}$, can reproduce the observed X-ray light-curve in the scenario put forward by K13 for 83~Ms. For considerably smaller $E$ and larger $A$, one or more of the conditions above are not satisfied.

To summarize, we found that the density of the wind environment must be still very small, while $E$ must still be substantially larger than $10^{53}$~erg. By adapting Eq. \ref{radius} to the $\rho (r) \propto r^{-1.4}$ case, we find that at 48~Ms after the trigger the ejecta are at $5\times10^{20}$~cm from the centre of the explosion, i.e. $\simeq160$~pc. We thus conclude that this model basically presents the same problems as those of P14.

In the model proposed by VA14, the density profile is again non-canonical, with $\rho (r) = A\,r^{-1.7}$. A few physical parameters of the two jets evolve in different fashion (see Tab. 3); moreover, the distance reached by the ejecta is rather unconstrained in the model of VA14: $R=(0.07-2)\times10^{19}~t_{\rm d}^{0.43}$~cm, where $t_{\rm d}$ is the time expressed in days. If we apply our lower limit of $48$~Ms for any change from a wind medium to a constant density medium, we have $R=3-100$~pc. Wind bubbles with radii towards the low end of this interval do not require an unusually small density of the pre-existing environment. The energy budget predicted by this model is uncertain as well; for our lower limit of the jet break time, a total energy\footnote{Assuming the minimum kinetic energies of both the narrow and wide jets described in this modeling; see VA1{\bf 4} for more details.} of $\sim10^{51}$ erg would be enough. We conclude that the model of VA14, in which the ejecta are moving in a stellar wind that has a non-standard profile, could still explain our late X-ray data, but we are concerned that it may do so more by virtue of the indeterminacy of some of its parameters than by any particular merits of the physical scenario which it describes.



\subsection{Constant density medium}\label{constant}
M14 have assumed that GRB 130427A occurred in a constant density medium. In their scenario, a jet break occurred at $\simeq37$~ks, but the post jet break slope of the flux is not steep because the FS physical parameters evolve with time. The fractions of energy given to the radiating electrons and to magnetic fields -- $\epsilon_{\rm e}$ and $\epsilon_{\rm B}$ respectively -- increase with time, while the fraction $\xi$ of electrons accelerated decreases with time. M14 assume $\epsilon_{\rm e} (t) = 0.027 \times (t/0.8{\rm d})^{0.6}$, $\epsilon_{\rm B} (t) = 10^{-5} \times (t/0.8{\rm d})^{0.5}$ and $\xi(t) =1 \times (t/0.2{\rm d})^{-0.8}$. In the model proposed by M14, the $\alpha=1.31$ detected in our observations should basically be a post-jet break decay, moderated by this evolution of microphysical parameters.

However, the data presented in our paper cover a much longer duration (by a factor of $\simeq20$) than those presented in M14. This duration of the temporal slope may make excessive demands on the scenario of evolving parameters. With the reasonable assumption that the maximum $\epsilon_{\rm e} = 1/3$ at equipartition, then this parameter saturates at $\simeq52~{\rm d} = 4.5$~Ms. Moreover, we would have $\xi\simeq10^{-3}$ by the time of the end of our observations; it is difficult to explain why such a tiny fraction of electrons are accelerated. Thus, $\epsilon_{\rm e}$ and likely $\xi$ cannot contribute to keep the post jet break decay slope less steep than expected. As a consequence, the fact that we see no steepening of the light-curve over 83~Ms (or at least until 61 Ms) weakens the ISM scenario under the assumption of an early jet break and evolving microphysical parameters.

We have already demonstrated at the beginning of this section that the light-curve does not have the steep decay slope of the lateral spreading jet, at least up to 61~Ms. Non-sideways spreading jets have less steep decay indices; the FS model predicts $\alpha = 3p/4 = 1.73$ for $\nu_{\rm x} < \nu_{\rm c}$ and $p=2.3$ used by M14. This prediction still fails to match the observations. 

Overall, our new data indicate that the X-ray emission after 40 ks is difficult to reconcile with a FS with jetted expansion, regardless of whether lateral expansion is present or not, and parameters are evolving as proposed by M14. Such a jet break would significantly reduce the energy budget of GRB 130427A down to level a magnetar could plausibly produce, which is $\simeq 10^{53}$ erg \citep{metz15}. But as we rule out an early jet break as proposed by M14, and derive a lower limit that is three orders of magnitude later, our new data are also problematic for the hypothesis of a magnetar central engine \citep{usov1992}, as proposed by \citet{mazzali14}.


\subsubsection{Standard on-axis model}
Could a simple FS model in an ISM medium, with no jet break or evolving parameters, explain the late GRB 134027A X-ray light-curve? First of all, we have to find a satisfying relationship between the observed decay and spectral indices $\alpha$ and $\beta$ in the FS model. For $\beta=0.79\pm0.03$, assuming spherical expansion, we should either have $\alpha = (3\beta+5)/8 = 0.92 \pm0.01$ for $\nu_{\rm c} < \nu_{\rm x}$ or $\alpha=3\beta/2 = 1.19 \pm0.05$ for  $\nu_{\rm c} > \nu_{\rm x}$. The former is clearly ruled out, but the latter is acceptable at $\simeq2.4 \sigma$ C.L..
If no jet break is present in our X-ray light-curve before 61 Ms, as our analysis indicates, then the ISM scenario predicts rather unusual values of the physical parameters involved. Let us first consider the total energy corrected by beaming. By definition, this parameter is

\begin{equation}
E_{\rm tot, corr} = (E_{\gamma,{\rm iso}} + E_{{\rm K,iso}}) f_b
\end{equation}
where $E_{\gamma,{\rm iso}}$ and $E_{{\rm K,iso}}$ are the energy emitted in gamma-rays and the relativistic kinetic energy of the ejecta respectively assuming isotropic emission, and $f_b \simeq \theta_{\rm jet}^2 /2$ is the beaming factor. Given the definition of the efficiency $\eta$ in converting the initial energy into prompt $\gamma$-ray photons (see Sect. \ref{wind_section}), one derives

\begin{equation}\label{b}
E_{\rm{tot,corr}} = \eta^{-1} E_{\gamma,{\rm iso}} f_b
\end{equation}

Following \cite{zha09}, the beaming angle $\theta_{\rm jet}$ in a constant density medium is

\begin{equation}\label{c}
\theta_{\rm jet} = 0.12 \left(\frac{t_{{\rm jet,d}}}{1+z}\right)^{3/8} \left(\frac{E_{{\rm K,53,iso}}}{n}\right)^{-1/8}\, \rm{rad}
\end{equation} 
where $n$ is the density of the medium in which the ejecta are expanding. The above formula implies that $f_b \propto E_{\rm k,iso}^{-1/4} n^{1/4}$. Remembering the definition of $\eta$, we find that $f_b \propto (\frac{1-\eta}{\eta})^{-1/4} E_{\gamma, {\rm iso}}^{-1/4} n^{1/4}$. Finally, substituting this last equation into Eq. \ref{b}, we derive that $E_{{\rm tot,corr}} \propto (\eta^3 -\eta^4)^{-1/4} n^{1/4} E_{\gamma, {\rm iso}}^{3/4}$. For any given $n$ and $E_{\gamma, {\rm iso}}$, the minimum $E_{{\rm tot, corr}}$ is obtained for $\eta = 3/4$. This efficiency value is high but it is not unprecedented, and models that entail magnetic dissipation (for example, see \citealp{zhang2011}) may explain it. In our case, $\eta = 0.75$ implies $E_{\rm{K,iso}} = \frac{(1-\eta)}{\eta} E_{\gamma,{\rm iso}}  = 1/3 \times 8.5\times10^{53}$~erg = $2.83\times10^{53}$~erg.

Detailed afterglow modeling of any long-duration GRB has not led to densities $n \stackrel{<}{\sim}$ a few $\times10^{-4}$ \citep{cenko2011,pkr02} for an ISM-like medium. Assuming $n=10^{-3}$, $\eta=3/4$, and the lower limit of $61$~Ms for the jet break time, we find that the minimum, beaming-corrected total energy associated with GRB 130427A is $E_{\rm tot,corr} = 1.23\times10^{53}$~erg, for a beaming angle of $\theta_{\rm jet} = 0.47$~rad. 

If the outflow decelerated at $\simeq 20$~s after the trigger (as appears to be the case from the early LAT and afterglow optical light-curves, see Ackermann et al. 2014) in such a low density environment, it would require (Eq. 1 of \citealp{mol07}) a very high initial Lorentz factor $\Gamma_0 \gtsim 1400$. Such a value for this parameter has not been observationally determined before in long GRBs, and it is $\sim3$ times as large as the typical initial Lorentz factor found in GRB modeling \citep{oat09}\footnote{Note that \citet{oat09} calculate the Lorentz factor at the forward shock onset, when the outflow is substantially decelerated. The initial Lorentz factor is about twice the value \cite{oat09} calculate.}.
For more typical densities of GRB environments, $n=0.1-1$~cm$^{-3}$, one would find a more mundane value of $\Gamma_0 \stackrel{>}{\sim} 600 - 800$, but the lower limit on the total energy increases up to $E_{{\rm tot,corr}} \simeq (4-7)\times10^{53}$~erg. Already the value of $E_{\rm tot,corr} = 1.23\times10^{53}$~erg inferred for $n=10^{-3}$ appears to be too close to or above the maximum energy that a magnetar central engine can produce, which is $\simeq 10^{53}$ erg. Instead, we think that the levels of energy required to power GRB 130427A in the ISM scenario may be explained by a black hole (BH) central engine, as we show in the following.

If the the relativistic outflow is produced by the Blandford-Znajek mechanism (BZ, \citealp{blz76}), the energy source to tap is the BH rotational energy, which is $E_{\rm{rot}} = 1/2 M_{\rm BH} c^2 f(a)$, where $f$ is a function of the rotational parameter $a= J c / M_{\rm BH} c^2$. For a very fast rotating hole with $a\simeq0.9$, we have $f\simeq0.30$. Even optimistic estimates show that the BZ mechanism can only convert up to $\simeq10\%$ of such a reservoir into kinetic energy of the jets \citep{zhang01,mck05,lee00}. Thus, the BH central engine should have a minimum mass of $\simeq 5 M_{\odot}$ (e.g. \citealp{komissarov09}). Such value is not implausible; in reality, however, one should expect a black hole in the range of 8-12 solar masses for more realistic values of efficiency of conversion. If the jets are instead powered by neutrino-antineutrino annihilation, the predicted efficiency is even lower. In such a scenario, assuming  that $\simeq 3 M_{\odot}$ accretes onto the BH, the energy available to power the jets is $\simeq3\times10^{52}$~erg or less (see \citealp{cenko2011} and references therein). We note that our estimate of $E_{\rm tot}$ does not include the energy that the ``central engine" of the explosion directed into other channels, for example the kinetic energy of the ejecta of SN~2013cq $E_{{\rm K,SN}}$, which can be as high as $\simeq 6.4\times10^{52}$~erg \citep{xu13,cano15}, X-ray flares and radiative losses, which would make the required total energy produced by the central engine even bigger. The viability of constant density scenarios should also be tested against observations presented in bands other than the X-ray, for example the radio one (see P14 for a discussion on this point).

As a side note, we point out that the lack of change of slope in the X-ray light-curve implies also that the ejecta have not piled up enough circumburst medium to be decelerated to sub-relativistic speed; in other words, the ejecta have not entered the so-called ``non-relativistic" (NR) expansion. If this transition had occurred, we would have seen a decay slope $\alpha_{\rm NR} = 1.77$ for $\beta=0.79$, constant density medium and $\nu_{\rm X} < \nu_{\rm c}$ \citep{gao13}. The transition to NR expansion is expected to occur at $t_{\rm NR}  = 970 (1+z) ( E_{\rm K,iso,53} / n ) ^{1/3}$ days \citep{zha09}.
Fitting the X-ray light-curve with a broken power-law which has a post-break slope $\alpha = 1.77$, we find a 95\% C.L. lower limit on such a break of $39$ ~Ms (i.e. $\simeq 450$~d). Given the formula above, we can set upper limits on the density of the environment, which must have been $n  \stackrel{<}{\sim} 70$ cm$^{-3}$. If we assumed that the transition has not occurred until the end our observational campaign, then $n \stackrel{<}{\sim} 7$ cm$^{-3}$. Under the assumption of a constant density environment, these results imply that GRB 130427A did not occur in a giant molecular cloud, which have typical densities of $10^3$ particles cm$^{-3}$ \citep{bergin2007}.


\subsubsection{Off-axis model}
The calculations above take the simplifying assumption that the observer is placed on the symmetry axis of the jet. Instead, the observer may be placed off-axis, as M14 themselves assume. Such a solution has the advantages of postponing the jet break time with respect to the on-axis observer and reducing the required energetics.
The light-curve will become substantially steeper only when the observer will see emission from the far side of the jet \citep{vdh13}, that is, when $\Gamma^{-1} \simeq \theta_{\rm jet} + \theta_{\rm obs}$ where $\theta_{\rm obs}$ is the angle between the observer and the jet axis. If the observer were on-axis, the jet break would occur when $\Gamma^{-1} \simeq \theta_{\rm jet}$.
Incorrectly assuming that the observer is placed on axis, while actually being off-axis, may cause to over-estimate the real beaming angle and thus the total energy. On the other hand, if the observer is largely off-axis, the received flux will be considerably lower. For example, from Fig. 9 of \cite{vdh13}, one can see that the pre-jet break flux is reduced by a factor of $\sim2.5$ if the observer is on the edge of the jet, i.e. $\theta_{\rm obs} = \theta_{\rm jet}$. In the even more extreme case $\theta_{\rm obs} > \theta_{\rm jet}$, the observer would basically detect a much weaker X-ray rich GRB or an X-ray flash \citep{heise03} rather than the bright and $\gamma$-ray-rich GRB 130427A. 

Using the formulation and results of \citet{vdh13}, we derived that for $\theta_{\rm obs} = 0.4~\theta_{\rm jet}$ the observed afterglow flux is diminished by a factor 1.75 with respect to the case in which $\theta_{\rm obs} =0$. To compensate for this reduction, we would require $E_{K, iso}$ to increase by a factor of 1.75 as well. The real $\theta_{\rm jet}$ would then be $\simeq \frac{1}{1+0.4} 1.75^{-1/8}$ the value calculated by Eq. \ref{c} if the observer were on-axis. With a density $n=10^{-3}$ cm$^{-3}$, we infer that the semi-opening angle of the ejecta of GRB 130427A would be $\theta_{\rm jet}  \gtsim 0.31$~rad. With $E_{\gamma,{\rm iso}} = 8.5\times10^{53}$~erg, the beam{\bf ing}-corrected total energy $E_{\rm tot, corr} \gtsim 6.5\times10^{52}$~erg. This is about half the value we find in the on-axis model. The initial Lorentz factor would increase only slightly (by a factor $\simeq1.07$) from the value determined for the on-axis observer. Moreover, the required efficiency is lower than 0.75, because $E_{\rm K}$ is now higher; we derive $\eta = 0.63$. To summarize, this off-axis scenario slightly eases the problems of the high energetics and the high efficiency of the on-axis case, but it needs $\Gamma_0$ to increase moderately with respect {\bf to} that scenario, in which this parameter is already unusually high.

Note, however, that the energy estimate would still increase by a factor $\simeq1.3$ if no jet break occurred over 83~Ms.

\subsubsection{Structured jet}
Another possibility we explored, because it might lead to a reasonable energy budget, is the so-called ``structured jet" \citep{meszaros98,zhang02,rossi02,panaitescu05}. In this case, the ejecta kinetic energy per solid angle $\epsilon = dE / d\Omega$ is not constant, but depends on the distance from the jet axis. We parametrize this density as 


\begin{equation}\label{d}
\epsilon (\theta < \theta_{\rm c}) = \epsilon_c~{\rm, } ~ \epsilon (\theta_{\rm jet}>  \theta > \theta_{\rm c}) \propto \theta^{\rm -k}
\end{equation}
where $\theta$ is the distance from the jet axis, and the constant density $\epsilon_{\rm c}$ below a certain opening angle $\theta_{\rm c}$ is introduced to avoid divergence. Thus the jet has a bright ``core", where the energy is the highest, and less energy in the ``wings". GRB 130427A is a very energetic event, so we will assume that the observer is along the jet axis and sees the emission from the bright core (as in the case of another very luminous event, GRB 080319B, see \citealp{racusin08}). In such conditions, a light-curve break will occur when the Lorentz factor of the core ejecta will reach $\Gamma_{\rm c} \simeq \theta_{\rm c}^{-1}$, and the decay will become mildly steeper. We assume that this break occurs before or at 400~s, when the X-ray emission from the afterglow was still outshone by that of internal origin, characterized by rapid variability. We calculated $\theta_c$ using Eq. \ref{c} but replac{\bf ed} $\theta_{\rm jet}$ with $\theta_{\rm c}$, and we assumed a low density $n=10^{-3}$ cm$^{-3}$ and efficiency of $\eta = 0.75$, because these values already led to smaller total energetics in the case of a homogenous jet. With these parameters, we derive $\theta_{\rm c} = 5.3\times10^{-3}$~rad (less than half a degree). The successive X-ray light-curve, with a decay slope of $\alpha=1.31$, can be explained by the structured jet. In this case, the decay index will depend on $p$ and $k$. To have $p>2$ we must be in the case $\nu_{\rm x} < \nu_{\rm c}$; with such a condition we derive $p = (2\beta +1) = 2.58$. Using Eq.~8 of \cite{panaitescu05} we derive $k=0.23$.

A steeper break, with slope $\alpha \simeq p$, will be visible when $\Gamma_{\rm c} \simeq \theta_{\rm jet}^{-1}$. We have a lower limit of 61 Ms on any break with a decay slope of $\simeq 2$. Thus, we can compute a lower limit on the ratio between $\theta_{\rm c}$ and $\theta_{\rm jet}$. If the break is at 400~s, then $\theta_{\rm jet} > (6.1\times10^7 / 400)^{(3/8)} \theta_{\rm c} \simeq 88~\theta_{\rm c} \simeq 0.47$~rad. 

Given that $\epsilon_{\rm c} = E_{{\rm k,iso}} / 4\pi = \frac{(1-\eta)}\eta~E_{\gamma,{\rm iso}}~/~12.56 = 2.3\times10^{52}$~erg sr$^{-1}$, we integrated Eq.~\ref{d} up to $\theta_{\rm jet}$ and derived the lower limit on the beamed-corrected kinetic energy of the structured jet $E_{\rm K,corr} = 2.1\times10^{52}$~erg. Under the assumption that the efficiency of the gamma-ray emission process was constant throughout the whole jet, the total energy is at least $E_{\rm tot,corr} = 8.5\times10^{52}$ erg. Note that such a value is the energy of a single jet; to compare it with the values obtained for the standard on-axis and off-axis jet in the previous subsections, we have to multiply it by two. Thus, we have a total energy of at least $1.69\times10^{53}$~erg; such a value is still rather high, as in the case of an on-axis and uniform jet. Furthermore, the initial Lorentz factor of the ejecta in the core region is still $\Gamma_{\rm c,0} \gtsim 1400$. The energy budget we have found should be multiplied by $\simeq1.3$ if we assume that no jet break occurr{\bf s} over 83 Ms, i.e. the whole time span of the observations, rather than 61 Ms.
We note that this model can accommodate the possible break at 37~ks determined by M14; this break might occur when the observer starts to receive emission from outside the core. In such a case, $\theta_{\rm c} \simeq 0.029$~rad (again assuming $\eta = 0.75$), which is closer to what is typically determined in the context of GRB modeling \citep{racusin09}. The lower limit on the ratio between $\theta_{\rm jet}$ and $\theta_{\rm c}$ decreases to 16.1; however, the beaming-corrected energy, the value of $\Gamma_{\rm c,0}$, and the opening angle lower limit are not different.

Finally, we point out that we have been quite conservative in deriving the lower limits on the beaming-corrected energy during our treatment of the models in {bf a} constant density medium. We have adopted 61~Ms as lower limit on the jet break time; such a value, however, is calculated assuming a post-jet break slope of 2.2 (see Sect. \ref{results}). We could have adopted a steeper slope of 2.58, derived from our finding $p=2.58$. Such a choice would have led to larger lower limits on a jet break time and thus on the beaming angle and, in turn, on the beaming-corrected energy. We preferred, however, to adopt a milder decay slope, which is more moderate and consistent with the modeling produced by other authors.







\subsection{Energy injection}
\cite{pan14} proposed that the outflow of GRB 130427A moves into a stellar wind medium; the X-ray band is below the cooling frequency $\nu_{\rm c}$, but the decay slope is not steep because of a process of energy injection \citep{zhang06}. This energy injection would be provided by Poynting flux-extracted energy produced by a newly born magnetar, or by shells catching up with the ejecta producing the emission. Another possible scenario is that a jet break occurs even before the beginning of observations, but the decay slope is shallower because the shocks are continuously refreshed \citep{schady07,depasquale10,troja12} by energy injection. This possibility can be applied to jets that are laterally spreading as well as to those which are not, and to both stellar wind and constant density medium scenarios. 

The observations presented in this paper are much more extensive than those analyzed by \cite{pan14}, and it is difficult to envisage a process of energy injection which could last for 83 Ms, i.e. almost 2 years in the rest frame of GRB 130427A. It has been argued that a magnetar central engine could power events such as some Ultra-Long GRBs (ULGRBs; \citealp{levan2014b}), such as GRBs 121027A and 111209A for timescales of several hours or a few days, and super-luminous SNe (SLSNe), for time scales of months or years \citep{kasen2010,greiner2015,metz15,cano2016,nicholl2013,inserra2013}. However, this scenario can likely be ruled out in the case of GRB 130427A. First, the magnetar is expected to inject energy into the ejecta at a steady pace, i.e. with a luminosity $L\propto t^0$. This is different from the injection energy rate hypothesised by \cite{pan14}, which is $L\propto t^{-0.3}$. Secondly, the shape of the spectrum of SN 2013cq \citep{xu13}, the SN associated with this event, is unlike those of SLSNe and even SNe associated with ULGRBs. On the other hand, if GRB 130427A is powered by a black hole, energy will be produced as long as an accretion disk exists. One typical timescale of the accretion is the free fall time of the matter of the progenitor envelope, which is expected to be few minutes. It might extend to a few hours if the progenitor is a blue supergiant, as postulated for the ULGRBs 111209A and 130925A (\citealp{gendre2013,piro2014}). However, this timescale obviously does not apply to the 80 Ms of the case at hand.
Another timescale is that of the so-called fall-back process, which is thought to occur in supernova explosions \citep{colgate1971}. Parts of the stellar envelope fail to reach the escape velocity, and end up falling onto the collapsed core of the exploded star. Such a fall-back process can last for tens of millions of seconds \citep{wong2014}. However, this fall-back process characteristically follows a $\dot{M} \propto t^{-5/3}$ law, where $\dot{M}$ is the mass accretion rate onto the central object. If the luminosity $L$ produced in this process is proportional to $\dot{M}$, then the hydrodynamics of the ejecta and the emission processes are not significantly affected, as shown by \citet{zhang02}. The energy injection process can change the afterglow light-curve only if its luminosity $L$ has a temporal behaviour $L\propto t^{-q}$ where $q<1$. 
The timescale for energy injection may be long if the ejecta form an accretion disk with low viscosity, and/or powerful magnetic fields slow down the accretion of matter onto the central object; nonetheless, the accretion time is not predicted to reach timescales of several tens of Ms \citep{kumar08}.

\vspace{1cm}

A summary of the key details of the models analyzed (e.g. energy, density profiles, physical parameters of the blast wave and emission mechanisms), along with very brief summaries of the problems that would arise due to the late time X-ray observations, are presented in Table~3~and~4.

\section{Conclusions}
In this article, we presented {\it XMM-Newton} and {\it Chandra} X-ray observations of the exceptionally bright {\it Swift} GRB 130427A, which have been carried out up to 83~Ms after the burst trigger. Such a timescale is unparalleled for X-ray afterglows of GRBs. We reconstructed the X-ray light-curve to very late epochs, and we find that the simple power-law decay with a slope of $\alpha = 1.309\pm0.007$, which was found from as early as ~30 ks after the trigger, extends up to the end of our observations. 
No jet break or flattening of the light-curve are visible. 

We discussed the consequences of this result with respect to the results of modelling presented in the literature, which considered data up to $\sim10$~Ms. We have treated the models in which the external medium is a stellar wind (P14, L13, VA14, K13) separately to that in which the medium has a constant density profile (M14). We find that the model of P14 requires a very low density of the pre-existing medium the GRB progenitor: $n_0 = $ few $\times10^{-4}$. Such a value is difficult to explain because the burst position is superimposed on the host galaxy and not external to it (where such low densities are more likely to occur), and there is non-negligible absorption at the redshift of the burst. The model of L13 does not require such a very low density, but instead it needs an extremely high efficiency to convert the initial energy into prompt $\gamma$-ray emission. The model put forward by VA14 considered several evolving and unconstrained parameters, which makes it difficult to test it against our data.  Finally, M14 assumed a constant density and a jet break, which is not steep because of evolving {\bf micro}physical parameters. We find it difficult that such a solution can keep a post jet-break decay slope as shallow as $\alpha\simeq1.3$ up to 83~Ms after the trigger. 

We have found that a{\bf n}  ISM scenario, in which the observer is placed on the jetted outflow axis, requires an exceedingly high energetics; a structured jet does not ease the problem. However, we propose that the ISM scenario, with an observer placed at $\theta_{\rm obs} = 0.4~\theta_{\rm jet}$, could still explain our observations with unusual physical variables: a beaming-corrected energy $E\simeq6.5\times10^{52}$ erg, an initial Lorentz factor of the ejecta $\Gamma_0 \gtsim1500$, an efficiency of~$\eta=0.63$, {\bf and} medium density $n = 10^{-3}$ cm$^{-3}$. Moreover, the central engine of GRB 130427A should be a black hole of a few solar masses which, by means of a very efficient Blandford-Znajek mechanism, should produce jets with opening angle $\theta\simeq0.31$~rad.

To summarize, the late X-ray behaviour of GRB 130427A, presented in this work, challenges external shock models discussed by previous authors. These models require extreme values of the physical parameters of the explosion, the emission mechanism and the environment.
We have found the least problematic scenario to be an off-axis jet in a constant density medium. Even this model, however, needs atypical parameters.

Further X-ray observations of GRB 130427A in 2016 can push the limit of energetics and/or density of the environment further and, in doing so, test the proposed models even more stringently. If the X-ray afterglow of GRB 130427A  continues to decay with the slope of $\simeq1.3$, its flux at the end of 2016 (roughly 115~Ms after the trigger) should be $\simeq2\times10^{-15}$ erg cm$^{-2}$ s$^{-1}$, which corresponds to $\simeq0.13$ counts ks$^{-1}$ with the ACIS-S camera onboard {\it Chandra}. A $\simeq100$~ks observation with this instrument would allow a detection of such a source and measure its flux with a $\simeq30\%$ precision.

The  {\it XMM-Newton} and {\it Chandra} X-ray observations of GRB 130427A, presented in this paper, constitute an important legacy for the study of this once in a lifetime event. These data have enabled us to explore the physics of GRBs in unexplored epochs, and highlighted problems with the standard models which will need addressing in the future. We showed that late time observations of bright GRBs are a powerful test of the models, and the sensitivity of facilities such as {\it Athena} \citep{nandra2013} will enable us to push the limits even further. In fact, even if the break in the light-curve should occur now, the X-ray flux expected at the time of the launch of \textit{Athena} (2028) is around $10^{-16}$ erg cm$^{-2}$ s$^{-1}$, thus being detectable by \textit{Athena} itself -- allowing to extend the time coverage by almost one order of magnitude more.

\begin{small}
\begin{table*}
\begin{center}
\begin{tabular}{ll}
\hline
Model \& Description & Problems \\ 
\hline
 & \\
Scenarios with $r^{-2}$ density profile                &   \\
													    & \\              											
Perley et al. 2014 (P14) 							   & The radius of the stellar bubble must be very large, $r \gtsim 105$~pc ($180$~pc). The required \\
$\theta_{\rm jet} \gtsim 0.09$ (0.10),                  & density of the pre- existing medium, in which the stellar wind bubble has been blown is  \\ 	
$E_{\rm tot,corr} \gtsim 5.1\times10^{51}$  ($5.9\times10^{51}$)~erg   & $n_0 \ltsim 2\times10^{-4}$ $(10^{-4})$. The value of this parameter is very low and unlikely in star  \\  
$A = 3\times10^{-3}$			                                &forming regions. \\
													    & \\
Laskar et al. 2013 (L13) 							    & The radius of the stellar bubble is smaller than in P14, and the density of the pre-existing  \\
$\theta_{\rm jet} \gtsim 0.13$ (0.15),                   & medium can be  higher and more realistic. However, this smaller radius is attained by   \\
$E_{\rm tot,corr} \gtsim 8.4\times10^{51}$  ($9.8\times10^{51}$)~erg  & assuming a much lower kinetic energy of the ejecta $7\times10^{52}$. In turn, this can be achieved only   \\
$A = 3\times10^{-3}$	                                           &  assuming a puzzlingly high radiative efficiency $\eta = 92\%$. \\
													    & \\
Panaitescu et al. 2013		    						& The substantial energy injection should carry on for tens of millions of seconds. The SN  \\ 			$E_{\rm K,iso} = 10^{54} \times (t/20ks)^{0.3}$		    & associated with GRB 130427A does not look like a magnetar-driven SLSN, which are events  \\
													    & possibly powered by a magnetar for such extended periods; in addition the energy injection \\
													    & from a magnetar is expected to produce a different X-ray afterglow light-curve. \\
													    & The fall-back mechanism in SNe can last tens of millions of seconds, but it would have a \\
													    & luminosity $L \propto t^{-5/3}$,\ which is much milder than that postulated in this model.\\
													    & Accretion power from a disk surrounding a stellar black hole is not predicted to last\\
												          & long enough. \\ 
												           &  \\															    
\hline
& \\											    
Scenarios with non-$r^{-2}$ profile 				     & \\
																     & \\													
Kouveliotou et al. 2013 (K13) 							           & The radius of the stellar wind bubble is again very large, $r\gtsim 160$pc, while the wind is still  \\
$E_{\rm iso} \simeq 1.8\times10^{54}$ ~erg; $\rho = 6\times10^{20}~r^{-1.4}$  & thin. As a consequence, the pre-existing density must have been extremely low, as in \\
					      										        &  the case of P14. \\
 																	  & \\
van der Horst et al. 2014 (VA14) 											& This model could account for the late X-ray data, but one concern is that it achieves this \\
$\rho \propto r^{-1.7}$ 												      & by several indeterminacies of its parameters. \\
$E_{\rm K,corr,n} = 3\times10^{53} - 3\times10^{54}$~erg  				& \\   
$E_{\rm K,corr,w} = 8\times10^{51} - 3\times10^{54}$~erg 				& \\
$\epsilon_{\rm B,n} = 10^{-4} - 10$, $\epsilon_{\rm B,w} = 8\times10^{-3} - 3$ & \\
$\epsilon_{\rm e,n} = \lbrack 0.08 - 0.8 \rbrack \times t_{\rm d}^{-0.4} $, $\epsilon_{\rm e,w} = \lbrack 1 - 7 \rbrack \times t_{d}^{-0.2}$ & \\
$\Gamma_{\rm n}  = \lbrack 0.6 - 1\rbrack \times10^2 t_{\rm d} ^{-0.28}$, & \\
$\Gamma_{\rm w}  = \lbrack 2-8 \rbrack \times10^1 t_{\rm d}^{-0.28}$  & \\
& \\
\end{tabular}
\caption{Essential parameters of the stellar wind models proposed and analysed in this paper. Nomenclature: $E_{\rm K}$ and $E_{\rm tot}$ are the kinetic energy of the outflow and the total energy (kinetic + prompt emitted in $\gamma$-ray) respectively. The semi-opening angle of the jet is indicated as $\theta$ and is given in radians. The suffixes ``iso" and ``corr" indicate the energy assuming isotropy and after beaming correction. Density $\rho$ of the medium is $A\times10^{35}~r^{-2}$ (stellar wind) particles cm$^{-3}$, where $r$ is the radius from the centre of the explosion.  Suffixes ``w" and ``n" indicate the wide and narrow jets in the double-component model of VA14. The fraction of energy possessed by electrons and magnetic field are $\epsilon_{\rm e}$, $\epsilon_{\rm b}$ respectively. Finally, $t$ and $t_{\rm d}$ express time and time in units of days. Numbers between round parentheses indicate the limits on the parameters if one assumes no break throughout the whole duration of observations.}
\label{}
\end{center}
\end{table*}
\end{small}
                       
\clearpage                        
                        
                        \begin{small}
\begin{table*}
\begin{center}
\begin{tabular}{ll}
\hline
Model \& Description & Problems \\ 
 \hline

& \\
Maselli et al. 2014 (M14) 													& The parameter $\epsilon_{\rm e}$ reaches its maximum equipartition value of 1/3, at 53~days; \\
$\theta_{\rm jet} = 0.059$													& however our observations show an interrupted power-law decay at least until $\simeq 700$~d.\\
$\epsilon_{\rm e} = 0.027 \times (t/0.8d)^{0.6}$;  $\epsilon_{\rm B} = 10^{-5} \times (t/0.8d)^{0.5}$  & Similarly, the fraction $\xi$ of accelerated electrons would be as low as $0.001$ at the end \\
$\xi =1 \times (t/2d)^{-0.8}$  														   & of our observations; it is not clear why this fraction should be so low.\\
																				   
																	& \\
On-axis jet model											     & Very large energetics and Lorentz factor -  one of the highest inferred so far for GRB  \\
$E_{\rm K,iso} = 2.83 \times10^{53}$~erg						& modeling. \\
$\theta_{\rm jet} \gtsim 0.47$										     & \\
$E_{\rm tot,corr} \gtsim 1.23 \times10^{53}$~erg						& \\
$n = 10^{-3}$, $\Gamma_{0} \simeq 1400$						& \\
																		& \\
Off-axis jet model											     & Lower energetics than in the case of on-axis model; however the Lorentz factor is even  \\
$E_{\rm K,iso} = 2.4\times10^{52}$~erg								     & higher. \\ 
$\theta_{jet} \gtsim 0.31$										     & \\
$E_{\rm tot,corr} \gtsim 6.5 \times10^{52}$						& \\
$n = 10^{-3}$, $\Gamma_{0} \simeq 1500$						& \\
       																	& \\
Structured jet model       															& The energetics are even larger than in the case of on-axis jet with no structure, \\
$\theta_{\rm c}, \theta_{\rm jet} = 0.0053, \gtsim 0.47$ (jet break at 400 s)	& while the Lorentz factor is not different. \\ 
$\theta_{\rm c}, \theta_{\rm jet} = 0.029, \gtsim 0.47$ (jet break at 37 ks)	& \\
$\epsilon (\theta_{\rm j} > \theta_{\rm c}) \propto \theta^{\rm -0.23}$ 		& \\
$E_{\rm K,corr} \gtsim 2.1\times10^{52}$~erg											& \\
$E_{\rm tot,corr} \gtsim 1.69 \times10^{53}$~erg											& \\
$n = 10^{-3}$, $\Gamma_{\rm 0, c} \simeq 1400$											& \\

\end{tabular}
\caption{Essential parameters of the models in a constant density medium proposed and analysed in this paper. Nomenclature is as in Table 3, except the density $n$ is in units of particles cm$^{-3}$, $\xi$ is the fraction of shock-accelerated electrons, $\Gamma_0$ indicates the value of the pre-deceleration Lorentz factor assuming no structure in the ejecta, $\theta_{\rm c}$, $\Gamma_{\rm 0,c}$ are the angular size and Lorentz factor of the core region of the ejecta assuming that they have a structure. In this case, $\epsilon$ represents the density of energy over solid angle.}
\label{}
\end{center}
\end{table*}
\end{small}

\clearpage

\section*{Acknowledgments}

MDP thanks Lara Nava and Daisuke Kawata for helpful discussions and references, Alice Breeveld for her helping hand, and his friends Marco D'Alessandro, John Moore, Ernesto Amato, Peter Rockhill, Peter V\v{c}asn\'y, Alexander J. Zech, and Pierluigi Cox for their encouragement.
AR acknowledge support from PRIN-INAF 2012/13 and from Premiale LBT 2013.
BG acknowledges financial support of the NASA through the NASA Award NNX13AD28A and the NASA Award NNX15AP95A. Part of this work is under the auspice of the FIGARONet collaborative network, supported by the Agence Nationale de la Recherche, program ANR-14-CE33.
DAK acknowledges support by TLS Tautenburg in form  of a research stipend.
DM acknowledges support from Ida as well as financial support from Instrument Center for Danish Astrophysics (IDA).
GS acknowledge the financial support of the Italian Ministry of Education, University and Research (MIUR) through grant FIRB 2012 RBFR12PM1F.
MDP and MJP acknowledge support from the UK Space Agency.
SRO acknowledges the support of the Spanish Ministry, Project Number AYA2012-39727-C03-01.
ZC gratefully acknowledges financial support from the Icelandic Research Fund (IRF).

This work made use of data supplied by the UK Swift Science Data Centre at the University of Leicester, and by the {\it Chandra} Data Archive.

\clearpage


\label{lastpage}

\end{document}